\documentclass[prl,twocolumn,amsmath,amssymb, a4paper]{revtex4}
\usepackage{natbib}
\usepackage{dcolumn}
\usepackage{graphicx}
\usepackage{bm}

\begin{document}

\bibliographystyle{apsrev}

\title{Dynamical Arrest in Attractive Colloids: The Effect of Long-Range Repulsion}
\author{Andrew I. Campbell}
\author{Valerie J. Anderson}
\author{Jeroen S. van Duijneveldt}
\author{Paul Bartlett}
\affiliation{School of Chemistry, University of Bristol, Bristol BS8
1TS, UK.}

\date{\today}

\begin{abstract}
We study gelation in suspensions of model colloidal particles with
short-ranged attractive and long-ranged repulsive interactions by
means of three-dimensional fluorescence confocal microscopy. At low
packing fractions, particles form stable equilibrium clusters. Upon
increasing the packing fraction the clusters grow in size and become
increasingly anisotropic until finally associating into a fully
connected network at gelation. We find a surprising order in the gel
structure.  Analysis of spatial and orientational correlations
reveals that the gel is composed of dense chains of particles
constructed from face-sharing tetrahedral clusters. Our findings
imply that dynamical arrest occurs via cluster growth and
association.
\end{abstract}

\maketitle

Systems far from equilibrium often exhibit complex structures even
while their equilibrium behavior may be quite mundane. A dramatic
example of this phenomenon is the behavior of suspensions of
attractive spherical particles which at equilibrium phase separate
but which, when concentrated rapidly, form a disordered arrested
solid -- a `gel' -- able to support a weak external stress
\cite{3373}. Although particle gels, formed from colloids or
concentrated protein solutions,  play an important role in many
areas of materials science as well as biology the molecular
mechanism of gelation remains far from understood \cite{faraday}.
Recently, in an attempt to understand the origin of the arrested
states which hinder protein crystallization, attention has focused
\cite{3357,3338,2545,2233,3276,3174} on gelation in systems where
short-range attractions (which favor particle aggregation) are
complemented by weak long-ranged repulsions (which provide a
stabilizing mechanism against gelation). Experiments
\cite{3357,3338,2545,2233} and simulation \cite{3276} reveal rather
unexpectedly a phase of stable, freely-diffusing clusters of
particles. The appearance of an equilibrium cluster fluid has lead
to suggestions \cite{3276,3174} that gelation in these systems
occurs via a cluster glass transition, where clusters (as opposed to
particles) are trapped within repulsive cages generated by the
long-range repulsion.

In this letter, we explore experimentally the effect of weak
long-range repulsive forces on dynamical arrest in a model system of
uniform colloids with short-range attractions. We use
three-dimensional fluorescence confocal microscopy to follow the
process of gelation directly with single particle resolution. In
sharp contrast with the structure of all gel networks studied to
date, we find that the connected solid network exhibits an high
degree of orientational ordering. Our findings suggest that kinetic
arrest occurs via a one-dimensional cluster growth and percolation
mechanism and not a cluster glass transition as has been proposed
previously \cite{3276,3174}.

\begin{figure*}[ht]
\includegraphics[width=5.5in,bb=22 63 560 245]{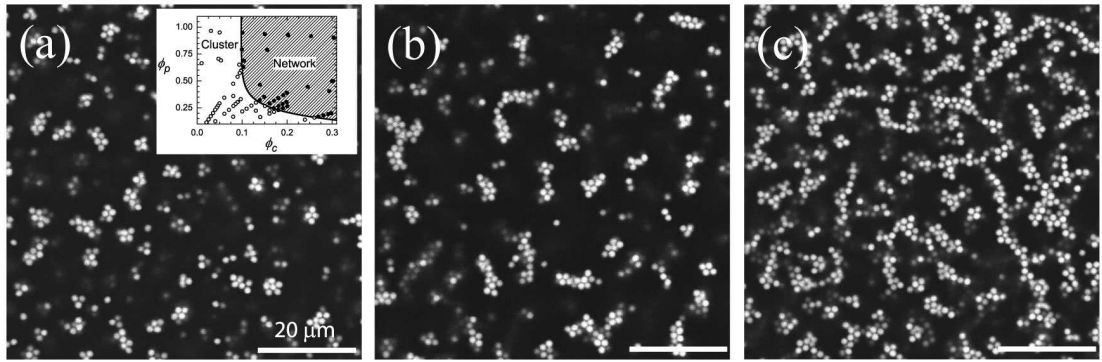}\\
\caption{Confocal microscope images of colloid-polymer mixtures at
different volume fractions. From left to right: $\phi_{c} = $ 0.080,
0.094, and 0.156. The attractive interactions are the same in all
samples, $U_{SR} = -9$k$_{B}$T ($\phi_{p} = 0.69$). Images (a) and
(b) contain clusters while (c) shows a network phase. The bars are
20 $\mu$m long. Inset: Phases observed as a function of the volume
fractions of colloid ($\phi_{c}$) and polymer
($\phi_{p}$).}\label{fig-images} \end{figure*}

Our particles are poly(methyl methacrylate) [PMMA] spheres
stabilized against aggregation  by a thin ($\sim10$nm)
covalently-bound layer of poly(12-hydroxystearic acid).  The spheres
have a mean radius of $a = 777$ nm, a polydispersity  of 3\% about
the mean, a mass density of $\rho_{c} = 1.176$ g cm$^{-3}$  and are
labelled with the fluorescent dye DiIC$_{18}$ \cite{2320}. The
spheres are dispersed in a mixture of cycloheptyl bromide (CHB) and
\textit{cis}-decalin, which nearly matches both the mass density and
the index of refraction of the colloid. In this mixture the PMMA
spheres have a small but reproducible positive charge. Phase
sensitive electrophoretic light scattering measurements reveal    a
net charge $Q$ of $+140e$ on each particle, equivalent to a surface
charge density $\sim 10^{-6}$ that of typical aqueous colloids.

The potential between spheres, $U(r)$, is the sum of long-ranged
screened coulombic repulsions $U_{LR}(r)$ and a short-ranged
attractive potential, $U_{SR}(r)$. The electrostatic repulsion has
the familiar Yukawa form, $U_{LR}(r) = (A/r) \mbox{ exp} [2 \kappa a
(1-r)]$. Here the Debye length $\kappa^{-1}$ reflects the effective
range of the coulombic repulsions screened by the ions in solution,
$r$ is the pair distance between particles scaled by the diameter
2$a$ and $A$ is the contact value of the potential. $\kappa^{-1}$ is
considerably extended by comparison to aqueous systems as a result
of the low solubility of ions. We estimate the ionic strength as
around $5 \times 10^{-9}$ mol dm$^{-3}$, corresponding to a Debye
screening length $\kappa^{-1} = 1 \pm 0.2 \mu$m \cite{ionic}. The
potential at contact, $A$, was estimated from electrophoretic
measurements as $+30k_{B}T$. To induce particle aggregation we
generate a short-range attraction between particles by adding
poly(styrene), a non-adsorbing polymer. We use a polymer with a mean
molecular weight of $10^{7}$ Daltons and a mean radius of gyration
$r_{g} \approx 92 $ nm. The attractive interactions induced occur at
pair separations less than $\xi = r_{g} / a$ or approximately 13\%
of the colloid diameter. Their strength is controlled by the polymer
concentration  and is equal to $U_{SR}/k_{B}T \approx -14 \phi_{p}$
at contact, assuming the polymer behaves ideally; here $\phi_{p}$ is
the fraction of the {\it free} volume of solution occupied by
polymer coils. Again, we emphasize the markedly different range of
the attractive and repulsive potentials; the attractions act on a
length scale $\xi a$ while the repulsions extend over a scale
$\kappa a \gg \xi a$.

To study the three-dimensional structure of particle gels we use
fluorescence confocal microscopy. The suspension was contained in a
cylindrical reservoir 10 mm in diameter and 650 $\mu$m deep created
by sealing the edges of a 170 $\mu$m coverslip to a microscope
slide. Fluorescence was excited by a He-Ne laser (543 nm) and imaged
through a 50 $\mu$m pinhole. The digital image was optimized using
the Nyquist criterion to select pixel size and frame spacing from
the instrumental resolution. The microscope's 63x  oil immersion
objective provided a field of view of 73 x 73 $\mu$m$^{2}$ with a
magnification of 71 nm/pixel. Typically, 345 vertically spaced
images were collected  at a separation of 0.16 $\mu$m.
Three-dimensional particle coordinates were extracted with an
accuracy of about 50 nm using algorithms similar to those described
in \cite{Crocker-966}.

Confocal microscopy images of colloid-polymer mixtures at fixed
attraction $U_{SR} = -9$ k$_{B}$T and three different particle
concentrations are shown in Fig.~\ref{fig-images}. At the lowest
volume fraction (Fig.~\ref{fig-images}a) the sample contains a fluid
of small, approximately spherical clusters. We confirmed that the
clusters were stable by repeatedly imaging the sample for over a
week, observing no significant growth in the size of clusters. With
increasing $\phi_{c}$ the clusters grow in extent, becoming less
spherical and more chain-like; the micrographs
(Fig.~\ref{fig-images}b) collected 4 hours after loading clearly
show short strands of densely packed chains of particles with a
relatively uniform thickness ($\sim $ 2$a$). The orientation and
center of mass of each strand fluctuates and the sample is ergodic.
Finally, at high densities ($\phi_{c} > 0.1$)
 the clusters form a connected solid network, which Fig.~\ref{fig-images}c reveals occurs with no change in local structure; the gel consists of linked densely-packed  chains of particles, reminiscent of percolation. Although the long-range structure of the network is highly disordered, there is a surprising degree of uniformity at short scales. The chains which form the network are clearly of nearly equal thickness. Repeated scanning reveals that while
individual chains display local fluctuations  all long-range motion
is suppressed and the sample is nonergodic. Using confocal
visualization we followed the transition between the ergodic cluster
fluid and nonergodic network as both the colloid density $\phi_{c}$
and the short-range attractions $U_{SR}$ were varied.  A
well-defined boundary between the two phases is clearly evident
(inset, Fig.~\ref{fig-images}) which we identify as the gelation
transition.

The spatial correlations evident in the gel are intriguing. The
order is in sharp contrast to all previous reports of colloidal gels
where the structure has been characterized locally either as a
scale-invariant fractal cluster (e.g. \cite{2545}) or, less
commonly, as an amorphous packing of hard spheres \cite{2598}. The
confocal images confirm that the local order is not a consequence of
gelation. The same local structure is clearly present in both the
fully ergodic cluster fluid (near the gelation boundary) as well as
the nonergodic gel. Indeed the main change at gelation is the degree
to which  chains are interconnected rather than any change in their
internal structure.  Thus we hypothesize that a weak long-range
repulsion coupled with a short-ranged attraction favors the
formation of dense strands of particles, the length of which grow
with increasing $\phi_{c}$ until, at the gelation transition, the
strands become completely interconnected and a gel forms. To test
this hypothesis we determine the three-dimensional coordinates of
our samples. We focus on the gels formed with attractive
interactions fixed at $U_{SR} = -13$ k$_{B}$T, referring to future
publications for a more extensive analysis.

\begin{figure}
  \includegraphics[width=3.2in,bb=20 20 190 75]{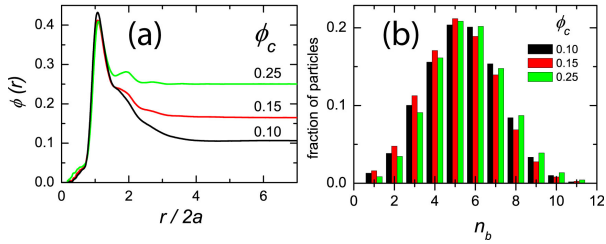}\\
  \caption{Correlations in colloidal gels as a function of $\phi_{c}$. (a) The local volume fraction $\phi(r)$
  and (b) the proportion of particles bonded to $n_{b}$ neighboring particles.
  Note that the peak in $\phi(r)$ at $r = 2a$ is independent of volume fraction and the
  distribution of particles bonds does not change with $\phi_{c}$.} \label{fig:phi-local}
\end{figure}

\begin{figure}
   \includegraphics[width=2.2in,bb=153 147 480 360]{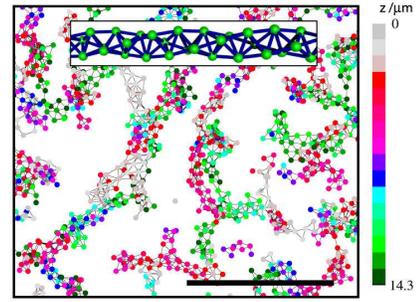}\\
  \caption{A two-dimensional projection of the particle centers within a slab of gel (14.3 $\mu$m deep) at $\phi_{c} = 0.1$.
   Particles are colored as a function of their depth within the sample and drawn 40\% of their actual size for clarity.
    The bar is 20 $\mu$m long. Inset: A spiral chain formed from tetrahedra of particles sharing faces.}\label{fig:marginal-gel}
\end{figure}

We first provide numerical evidence of correlations that are
independent of density and characteristic of the chains that form
the network. We start by analyzing the translational order present.
From the particle coordinates we compute the pair correlation
function $g(r)$ and hence the local density profile $\phi (r) =
\frac{24 \phi_{c}}{r^{3}} \int_{0}^{r} s^{2} g(s) ds$, which is the
average volume fraction within a sphere of radius $r$ centered on
each particle. Fig.~\ref{fig:phi-local}(a) shows $\phi(r)$ evaluated
for different gel densities. The local volume fraction decays
rapidly towards $\phi_{c}$ on distances of  3--4 particle diameters,
indicating that  translational order is short-ranged at all volume
fractions. The situation is very different at shorter scales ($r
\approx 2a$) where translational order is enhanced;
 on these scales Fig.~\ref{fig:phi-local}(a) indicates that the local
environment is dense and essentially independent of $\phi_{c}$; the
`cage' around each particle has a volume fraction $\sim 0.5$ (from
the peak of $\phi(r)$) and a size $\approx 2a$. To probe further the
local organization we examine the orientational order. We define
particles which are `bonded' to each other by their separation $r$.
If $r \leq r_{0}$, where $r_{0}$ is identified with the position of
the first minimum in $g(r)$, then the particles are considered
neighbors. This definition ensures that all particles in the first
coordination shell are counted as near neighbors.
Fig.~\ref{fig:phi-local}(b) shows the distribution in the number of
bonds per particle as the gel density is varied. The bond
distribution is essentially independent of $\phi_{c}$, confirming
the picture of the gel as a network of invariant chains. The mean
number of bonds per particle is  $n_{b} = 5.6 \pm 0.4$, midway
between the values of 2--3 expected for a fractal structure and the
12 bonds found in dense liquids and glasses. More dramatic evidence
for the three-dimensional order within the gel is shown in
Fig.~\ref{fig:marginal-gel}, which shows a two-dimensional
projection of particle centers and bonds identified within a small
slab of gel. The image reveals that the gel consists of relatively
uniform chains constructed from dense clusters of particles.

To check for local tetrahedral coordination we calculate the
rotational invariants $q_{l}(i)$, $\hat{w}_{l}(i)$ (for $l = 4$ and
6), which are quantitative measures of the local bond topology
around particle $i$. We define the bond-order parameters following
\cite{2850}. Structural and thermal fluctuations result in a
distribution of $q_{4}$, $q_{6}$ $\hat{w}_{4}$ and $\hat{w}_{6}$.
The resulting bond-order distributions provide a sensitive measure
of the symmetry of the local environment \cite{2850}. The
$\hat{w}_{6}$-distribution, in particular, facilitates the detection
of structures with tetrahedral symmetry since a large negative value
signals a high degree of tetrahedral order. A tetrahedral cluster
has $\hat{w}_{6} = -0.106$ while an icosahedron, an ordered
arrangement of twenty tetrahedra, maximizes $|\hat{w}_{6}|$ at
$\hat{w}_{6} =  -0.170$ \cite{Steinhardt-1290}. By contrast, the
cubic symmetries found in \textit{fcc}, \textit{hcp} and
\textit{bcc} crystals lead to near zero values of $\hat{w}_{6} =
-0.013$, -0.012, and 0.013 respectively.

\begin{figure*}
  \includegraphics[width=6.0in, bb = -5 -5 307 60]{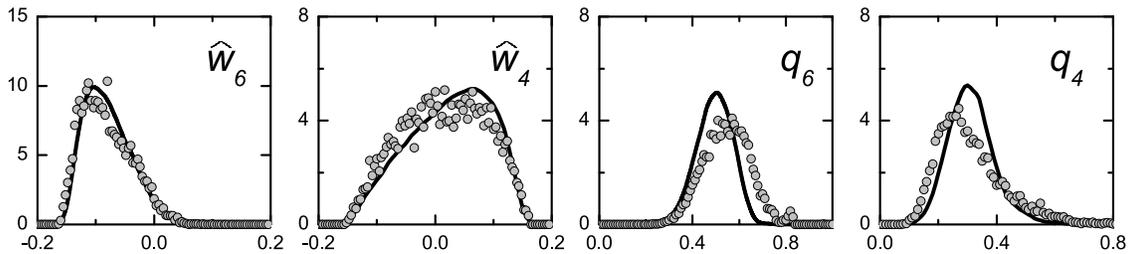}\\
  \caption{Orientational order in a colloidal gel with $\phi_{c} = 0.10$ and $U_{SR} = -13$k$_{B}$T.
  Measured $\hat{w}_{6}$, $\hat{w}_{4}$, $q_{6}$ and $q_{4}$ bond-order parameter distributions (points) together
  with values calculated from a model of face-sharing tetrahedra
  (black curves).}\label{fig:bond-order}
\end{figure*}

Figure~\ref{fig:bond-order} shows the bond-order histograms measured
for the gel at $\phi_{c} = 0.10$ (the distributions obtained at
other densities were similar but are not shown). The pronounced
maximum in the $\hat{w}_{6}$-distribution at $\hat{w}_{6} \sim
-0.13$ is particularly striking and demonstrates a high degree of
orientational order. The peak is at too low a value however to be
accounted for by tetrahedral units alone and the small number of
bonds per particle is incompatible with an icosahedral symmetry.
Although neither structure is consistent with the data, the position
of the $\hat{w}_{6}$ maximum, intermediate between the values
expected for tetrahedral and icosahedral clusters, suggests that the
gel probably contains \textit{ordered} clusters of tetrahedra.

Polytetrahedral order has been invoked to understand the structure
of quasicrystals, atomic liquids and glasses \cite{407,2869,2878}
however the concept has not been applied to gels, as far as we know.
A viable structure must account for the chains evident in the
confocal images together with the strong propensity for six-fold
coordination found experimentally. These constraints eliminate many
possibilities. So for example, while linear chains of edge- or
corner-sharing tetrahedra are common motifs in structural chemistry
the number of bonds per particle varies along the chain and averages
out at less than 6. More dense tetrahedral chains can be built  by
sharing \textit{faces} of neighboring tetrahedral units. There is a
single dimer of two face-sharing tetrahedra with the symmetry of a
trigonal bipyramid, one trimer (a bicapped tetrahedron), two
distinct tetramers and five pentamers. The number of such
arrangements grows rapidly as more tetrahedral units are added so
that an infinite sequence of face-sharing tetrahedra may exist in
any one of an infinite number of possible configurations, each with
a different chain contour. Each configuration however has an
identical shell of six nearest neighbors since the neighbors of each
particle belong to a trimer of adjacent tetrahedra, which may be
realized in only one way. Here we focus on the most striking
configuration, the graceful Bernal \cite{407} (or Boerdijk-Coxeter
\cite{2906,1}) spiral, reproduced in Fig.~\ref{fig:marginal-gel}. We
have calculated the distributions of bond-order parameters from
computer-generated coordinates for a spiral strand consisting of
1000 tetrahedra. Thermal and structural fluctuations  were simulated
by adding random deviations, chosen from a Gaussian distribution of
width $\Delta a$, to the ideal coordinates.
Fig.~\ref{fig:bond-order} depicts the bond-order distributions
calculated for the Bernal spiral with $\Delta$ = 0.21 where it is
clear the measured $\hat{w}_{6}$- and $\hat{w}_{4}$-histograms are
reproduced rather well by this model, strongly supporting our
picture of the gel as a network constructed from chains of
face-sharing tetrahedra. The measured $q_{6}$- and
$q_{4}$-distributions are less well reproduced but this is probably
linked to the fact that our model only has six-fold coordinated
species while the measurements contain particles with both more and
fewer than 6 nearest neighbors.

In summary, we have shown that the mechanism of dynamical arrest is
profoundly altered by the presence of long-range repulsive
interactions. Three-dimensional confocal microscopy reveals the
formation of stable chain-like clusters, constructed from
face-sharing tetrahedral units, which associate into a fully
connected network at gelation. The possibility that a delicate
balance of attraction and long-range repulsion may stabilize
equilibrium clusters has been discussed in recent theoretical work
\cite{3247,3267}. However it is not clear if the formation of
tetrahedra reflects the presence of charge or is simply a
consequence of a centro-symmetric attractive potential. An
explanation for the unexpected structures seen in our experiments
remains an open challenge to theorists. Finally it is tempting to
draw analogies between the one-dimensional growth seen here and the
tendency to chain formation observed in several protein solutions
\cite{3302,3253}. These similarities suggest that weakly-charged
colloidal systems could offer valuable new insights into the
microscopic mechanism of protein nucleation and gelation.

\begin{acknowledgments}

We thank Macolm Faers, Adele Donovan and Laura Starrs for extensive
discussions and help during the experiments. This work was supported
by EPSRC, Bayer Cropsciences and MCRTN-CT-2003-504712.

\end{acknowledgments}


\begin{thebibliography}{20}
\bibitem{ionic} The ionic strength was determined from the measured sample conductivity (170 pS cm$^{-1}$) assuming ion diffusivities of 6 and 2 $\times 10^{-10}$ m$^{2}$ s$^{-1}$, typical of Br$^{-}$ and alkyl
cation diffusion. The Debye length was calculated at $\phi_{c} =
0.1$, allowing for the screening from the solvent ions and the $Q$
counter-ions associated with each particle.
\end{thebibliography}
\end{document}